\documentclass[%
 reprint,
nofootinbib,
 amsmath,amssymb,
 aps,
floatfix,
]{revtex4-2}

\usepackage{graphicx}% Include figure files
\usepackage{dcolumn}% Align table columns on decimal point
\usepackage{bm}% bold math
\usepackage{physics}
\usepackage{subfigure}
\usepackage{xcolor}
\usepackage[hidelinks]{hyperref}
\usepackage[capitalize]{cleveref}
\usepackage{changes}

\begin{document}

\preprint{APS/123-QED}

\title{Control of Plasmons in Topological Insulators via Local Perturbations}% Force line breaks with \\

\author{Yuling Guan}
  \email{yulinggu@usc.edu}
 \affiliation{Department of Physics and Astronomy, University of Southern California, Los Angeles, CA 91361}
\author{Zhihao Jiang}
\affiliation{Department of Physics and Astronomy, University of Southern California, Los Angeles, CA 91361}
\author{Stephan Haas}
\affiliation{Department of Physics and Astronomy, University of Southern California, Los Angeles, CA 91361}

\date{\today}% It is always \today, today,

\begin{abstract}

 We use a fully quantum mechanical approach to demonstrate control of  plasmonic excitations in prototype models of topological insulators by molecule-scale perturbations. Strongly localized surface plasmons are present in the host systems, arising from the topologically non-trivial single-particle edge states. A numerical evaluation of the RPA equations for the perturbed systems reveals how the positions and the internal electronic structure of the added molecules affect the degeneracy of the locally confined collective excitations, i.e., shifting the plasmonic energies of the host system and changing their spatial charge density profile. In particular, we identify  conditions under which significant charge transfer from the host system to the added molecules occurs. Furthermore, the induced field energy density in the perturbed topological systems due to  external electric fields is determined.

\end{abstract}

\maketitle

%\tableofcontents

\section{INTRODUCTION}

Topological insulators (TIs) are characterized by a gapped bulk energy spectrum and symmetry-protected conducting edge states living on the surface. Prominent examples of the research on TIs include the discovery of integer quantum Hall effect\cite{kdp80,km05,km052,ztsk05,bz06,bhz06,kwbrbmqz07}, the two-dimensional (2D) quantum Hall insulator\cite{bf96,hk10,hp14}, which can host chiral edge currents, as well as three-dimensional (3D) TIs whose topological surface states are formed by massless Dirac fermions \cite{fkm07,hqwxhch08,caclmqzldfzfhs09}. TIs can also be realized in one spatial dimension, for example, in the paradigmatic example of the Su-Schrieffer-Heeger (SSH) model\cite{ssh79}. In the topologically non-trivial phase, the SSH chain has two localized single-electron edge states on its boundaries. Recently, collective excitations, such as plasmons, in the SSH chain and other TIs have been studied \cite{k11,om13,seth13,daywswz17,els12,jrgh20,sjh21}. Here, strongly localized plasmon modes were observed on the  boundaries of the SSH chain, which can be traced back to  the localized single-particle edge states.

One essential property of edge states in TIs is their robustness against disorder. For example,  quantum Hall currents are immune to back-scattering from any surface impurity\cite{ljs2014}. Single-electron topological states are well protected if their symmetry is preserved. However, whether the collective excitations in TIs are protected, completely or partially, is still an interesting and not fully resolved question. Unlike single-electron states, collective excitations are correlated phenomena routed in interactions\cite{dw17,sdh90,sgb13}. It has recently been shown that the plasmon edge modes in the SSH model are less robust against global hopping disorder than their constituent single-electron edge states\cite{jrgh20}, which is mainly due to the screening effect from the bulk electronic bands \cite{ssh79,aop16}. Impurities will occur in TIs\cite{llxqz09,hwz11,aay13}. In this work, we focus on the effects of \text{local impurities} on plasmonic excitations in the SSH model and a mirrored variant model (mSSH) connecting two SSH chains with distinct indices of topological invariant. Specifically, the local impurities here are modeled by diatomic molecules that can be placed at any position close to the unperturbed 1D host material. While these local inpurities are not expected to change the single-electron spectrum of the host material drastically due to topological protection, they can bring about other phenomena, such as electron tunneling and Coulomb coupling between the impurity molecule and the host material. As discussed below, we clearly observe such effects of local impurities on the localized plasmons excited at the boundaries (domain walls) of the 1D SSH (mSSH) model, which are strongly dependent on the positions of impurity molecules. As a reference point, we also study local impurities in proximity to a simple 1D metallic chain which hosts only extended plasmon modes propagating in the bulk. In both cases, topological and trivial, we discuss how control of  impurity positions can be  used as a tuning knob for manipulating the plasmonic excitations.

The remainder of this paper is organized in the following way. In Sec.~\ref{sec:2}, we introduce the real-space random phase approximation (RPA) method, which we use for all calculations of plasmons in this work. In Sec.~\ref{sec:3}, we focus on the three models under consideration: the 1D homogeneous metallic chain, the 1D SSH model, and the 1D mSSH model. We discuss in detail the effects of diatomic-molecule impurities on plasmonic excitations in these host models, with varying impurity positions and the internal coupling within the perturbing molecule. This is followed by a conclusion and outlook in Sec.~\ref{sec:4}

\section{METHOD}\label{sec:2}
In order to study the plasmonic excitations in TIs, we use the real-space random phase approximation (RPA) which has been introduced in previous studies \cite{pb52,bp53,bp51}. One major advantage of this approach is that we can directly obtain the real-space charge oscillation pattern for each single plasmon mode, as will be explained below. 

We first evaluate the non-interacting charge susceptibility function in the atomic basis via\cite{wvky18}
\begin{equation}
[\chi_0(\omega)]_{ab}=2\,\sum_{i,j}\frac{f(E_i)-f(E_j)}{E_i-E_j-\omega-i\gamma}\psi_{ia}^* \psi_{ib}\psi_{jb}^*\psi_{ja},
\end{equation}
where the independent variable $\omega$ is the excitation frequency in the unit of eV. We consider finite lattices with open boundaries. Therefore, $a$ and $b$ run over all atomic lattice sites. $E_i$ and $\psi_i$ are the electronic eigenenergy and eigenstate of the $i-$th level, which is here obtained by diagonalizing the model Hamiltonian. $f(\cdot)$ is the Fermi function, which in the following is approximated by a step function, namely, by its zero temperature limit. The factor 2 accounts for the spin degeneracy. $\gamma=0.01\ \text{eV}$ is a finite broadening. 
%\added{, and the constant damping $\gamma$ = 0.01 $eV$ describes the numerical broadening.}

Electron-electron Coulomb interactions are then introduced into the calculation on the RPA level. To do this, we evaluate the bare Coulomb interaction matrix $V_{ab}$ in the same atomic basis. To avoid a divergence, we use the Ohno potential\cite{pta04} with a proper cutoff parameter $\Delta$. The Coulomb interaction between two sites $\vec{r}_a$ and $\vec{r}_b$ is therefore given by  
\begin{equation}
\label{eq:Coulomb}
V_{ab}=
\frac{e^2}{4\pi\epsilon_\text{env}\sqrt{|\vec{r}_a-\vec{r}_b|^2+\Delta^2}}, \end{equation}
with $\epsilon_\text{env}$ being the dielectric constant of the background environment. In our calculation, we set $\Delta$ to be 1 \AA, which is smaller than the 1D lattice spacing $d = 3$ \AA. The real-space RPA dielectric response function (a matrix) is then  calculated from\cite{wcjtwm15}
\begin{equation}
\epsilon_\text{RPA}(\omega) = \text{I} - V\chi_0(\omega).
\end{equation}
We identify the plasmonic excitations from the electron energy loss spectrum (EELS), which is defined by \cite{wcjtwmn15,ajt12}
\begin{align}
    \text{EELS}(\omega)=\max_n \bigg \{ -\text{Im}\ \left[\frac{1}{\epsilon_n(\omega)}\right] \bigg \},
\end{align}
for each single frequency $\omega$, where $\epsilon_n$ is the $n$-th eigenvalue of $\epsilon_\text{RPA}(\omega)$. The EELS is peaked at plasmon frequencies. Letting $M$ be the selected index that maximizes $-\text{Im}[1/\epsilon_n(\omega_p)]$ at a plasmon frequency $\omega_p$, we can then simultaneously obtain the real-space charge distribution pattern of the corresponding plasmon mode $\omega_p$ by using the eigenstate $\psi_M(\omega_p)$ of the dielectric matrix, multiplied by the non-interacting susceptibility matrix $\chi_0(\omega_p)$\cite{wcjtwmn15}, namely,
\begin{equation}
\label{eq:ChargeDensity}
\rho_0(\omega_p)=\chi_0(\omega_p)\psi_{M}(\omega_p).
\end{equation}
Here, $\rho_0(\omega_p)$ is the induced charge density vector in the atomic basis, representing the plasmonic eigenmode at frequency $\omega_p$. 
Similarly, we can define the second EELS (2nd EELS) at any single frequency $\omega$ by selecting the second maximum of $-\text{Im}\ \left[\frac{1}{\epsilon_n(\omega)}\right]$ among all eigenvalues $\{\epsilon_n(\omega)\}$. A peak in the 2nd EELS indicates a degenerate mode (at least two-fold), which is  due to the symmetry of the underlying model. In fact, for a two-fold degenerate plasmon mode, the 1st EELS and the 2nd EELS coincide, indicating a degenerate subspace furnished by degenerate eigenvectors of the dielectric matrix $\epsilon_\text{RPA}(w)$. In general, one can define the third, forth, etc., EELS in the same manner. In this paper, we focus on 1D models, for which the 1st EELS and 2nd EELS are usually sufficient for the analysis.  

While the EELS yields the plasmonic eigenmodes of a system, it does not give any information of the system's response to a specific external field that is applied in experiment to excite the system. In this case, we need to calculate the induced charge distribution due to a specific external field $\phi_{ext}(\omega)$ via
\begin{equation}
\rho_{ind}(\omega)=\chi_\text{RPA}(\omega)\phi_{ext}(\omega),  
\end{equation}
using now the (interacting) RPA charge susceptibility function
\begin{equation}
\chi_\text{RPA}(\omega)=[\text{I} - \chi_0(\omega)V]^{-1}\chi_0(\omega).
\end{equation}
The induced potential, induced electric field, and the induced field energy density in real space can then be determined from $\rho_{ind}(\omega)$. By integrating the induced field energy density over the full space of consideration, we obtain the frequency-dependent induced energy spectrum, $U_{ind}(\omega)$, which also peaks at the plasmon frequencies ~\cite{jrgh20}.

\section{Results and Discussion}\label{sec:3}
\subsection{Plasmonic excitations in a decorated one-dimensional metallic chain - non-topological case}\label{sec:3.1}

\begin{figure}[h]
\centering
\begin{minipage}{\linewidth}
\centering
\includegraphics[width=\linewidth]{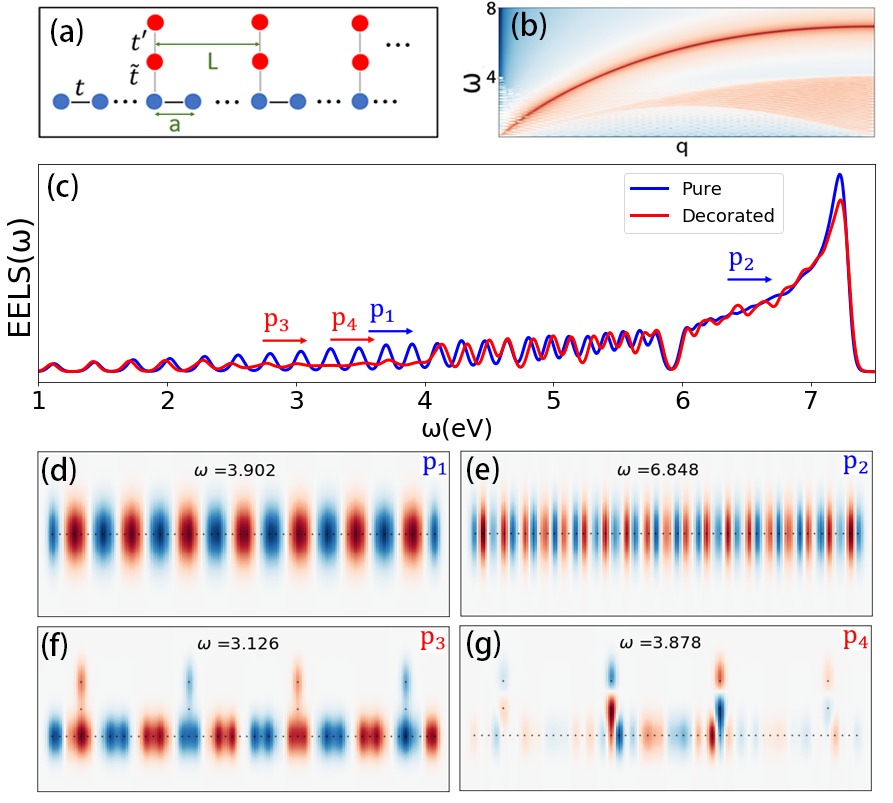}
\end{minipage}
\caption{(a) Illustration of a 54-site decorated metallic chain with hopping $t=1.0 \text{ eV}$, $\tilde{t}=0.5 \text{ eV}$ and $t'=2.0 \text{ eV}$ (b) Energy dispersion of a homogeneous metallic chain in momentum space. (c) Blue: EELS of a finite homogeneous metallic chain with open boundary conditions. Red: EELS of the same open-ended metallic chain with additional diatomic molecules like (a). (d)-(g) Real-space charge density modulation for (d) a low-energy plasmon in the pure metallic chain. (e) a high-energy plasmon in the pure metallic chain. (f) a low-energy plasmon in the decorated chain. (g) a confined low-energy plasmon in the bulk at $\omega=3.878$ eV of the decorated chain.}
\label{fig:Homo}
\end{figure}

Before discussing plasmons in topological insulators, let us first consider a homogeneous one-dimensional (1D) metallic chain (MC) with open boundaries as a benchmark, described by the real-space tight-binding Hamiltonian,   
\begin{equation}
\hat{H} = t \sum_{n=1}^{M-1}(\ket{n+1}\bra{n}+H.c.) + \mu \sum_{n=1}^{M}\ket{n}\bra{n},
\label{Hami_Homo}
\end{equation}
where $M$ is the number of the atoms in the chain, $t$ is the nearest-neighbor hopping parameter, and $\mu$ is the chemical potential. We explore the plasmonic excitations in this simple model by calculating the EELS for a finite chain wit $M=54$ sites, $t=1.0\text{ eV}$ and $\mu$ =-1.0 eV, which is shown by the blue line in Fig.~\ref{fig:Homo}(c). The EELS is made of a (quasi-)continuum of plasmonic excitations in the frequency range between $\omega = 0$ eV and  $\omega = 7.28\text{ eV}$. Furthermore, we observe a pseudo-gap at around  $\omega \approx 6\text{ eV}$, which arises from finite size effects.  For periodic boundary condition, this pseudo-gap is absent, and we have verified numerically that with increasing chain length it gradually disappears. Fig.~\ref{fig:Homo}(b) shows the momentum space dispersion of the pure metallic chain, which  corresponds to the pure EELS shown in Fig.~\ref{fig:Homo}(c). In the high energy region, the plasmon dispersion curve in momentum space becomes flat, leading to the van-Hove singularity observed at the maximum excitation energy. The low-energy plasmonic excitations arise from two-particle  processes combining distant electrons that are energetically close to the Fermi surface, whereas the high-energy plasmons arise from scattering electrons that are in close proximity to each other. 
%The plasmonic density of states (PDOS) is much larger at high frequencies than at low frequencies. [EXPLAIN WHY] We observe a continuous spectrum in the range $\omega \in \{6.039\text{ eV}, 7.28\text{ eV}\}$. 
All plasmons in this pure 1D MC are bulk modes. Two representative examples of different excitation energies are shown in Figs.~\ref{fig:Homo}(d) and (e) with their real-space charge modulation patterns plotted. We can see the low-energy mode [Fig.~\ref{fig:Homo}(d)] displays a longer wavelength, corresponding to slowly propagating waves. In contrast, the high-energy mode shows a much shorter wavelength, characterized by a rapidly oscillating charge modulation pattern. The enhanced phase space spectral density at high energies leads to the van Hove singularity, arising from two-particle scattering processes, connecting single electron states from the band minimum to those at the band maximum. 

Next, we investigate a decorated metallic chain (DMC) by adding four diatomic molecules with equal spacing to the homogeneous host system, as illustrated in  Fig.~\ref{fig:Homo}(a). Specifically, these four molecules are located above sites $4, 19, 34\text{ and }49$ of the  open-ended $M=54$-site host chain, respectively. They are aligned vertically, with an internal hopping $t'=2.0$ eV between the two atoms of the molecule, and a small tunneling hopping $\tilde{t}=0.5$ eV between the molecule and the chain. The red line in Fig.~\ref{fig:Homo}(c) shows the corresponding EELS of the DMC. Compared with the EELS of the pure MC, we see that for frequencies below $\approx 2\text{ eV}$ and above the pseudo-gap at $\approx 6\text{ eV}$, the spectrum of DMC is almost unaffected by the decorating molecules. In the intermediate frequency regime between $4\text{ eV}$ to $6\text{ eV}$, the spectrum structure of the pure MC is approximately preserved, with slightly shifted energies. In this regime, molecules are far off resonance. There is little charge tunneling between the molecules and the host MC. However, the spectrum within the energy window  $\omega \in \{2.7\text{ eV}, 4\text{ eV}\}$ is strongly affected due to the added molecules. Moreover, the molecules are on resonance and are excited as well, leading to significant charge transfer due to the interactions between the decorating molecules and the host 1D MC (see e.g. Fig.~\ref{fig:Homo}(f)). Furthermore, a pattern of oscillating charges between nearby decorating molecules is observed. In Fig.~\ref{fig:Homo}(g), we find another interesting mode, where the charge density in the 1D MC is confined due to the addition of the molecules. These two modes are not observed in the unperturbed host MC, and they both occur in the most affected energy regime for the chosen parameter set. 
\label{sec:A}

\subsection{Plasmonic excitation in the SSH model}\label{sec:3.2}
\begin{figure*}[t]
\centering
\includegraphics[width=\linewidth]{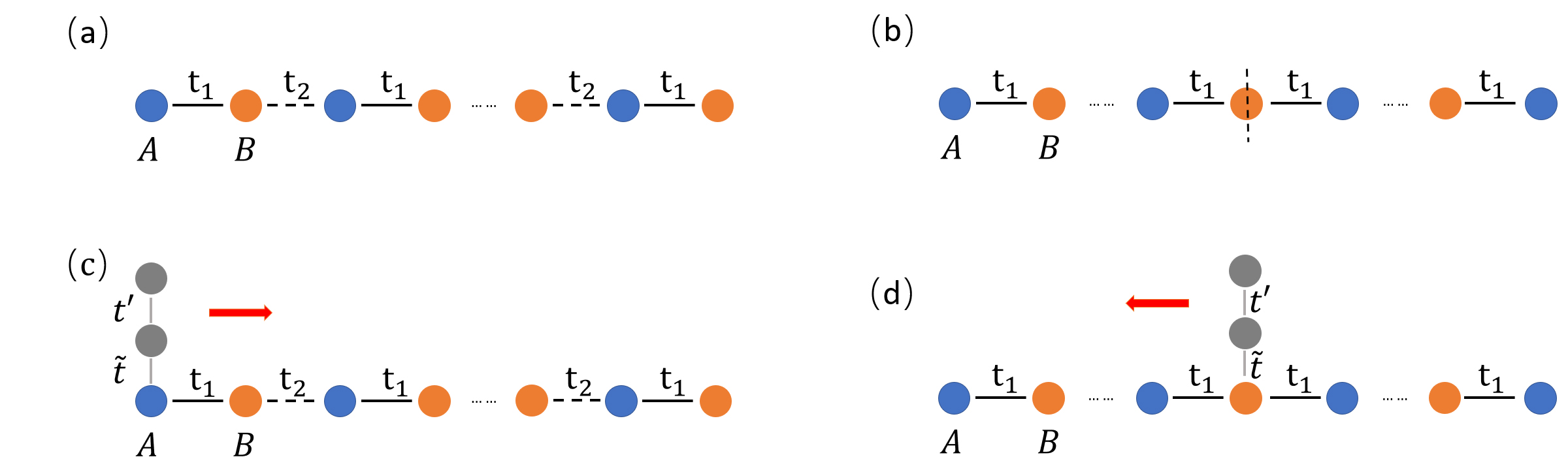}
\centering
\caption{(a) Illustration of the SSH model on a bipartite tight-binding chain. (b) Illustration of the mirror SSH model on the tight-binding chain with a mirror inversion at its center. (c) Illustration of the SSH model with an additional diatomic molecule, connected with one site of the chain. The position of the connected site is varied from the edge to the bulk. (d) Illustration of the mirror SSH model with an additional diatomic molecule,  connected with one site of the chain. The position of the connected site is varied from the center of the bulk to the edge.}
\label{fig:models}
\end{figure*}

Let us now turn to a prototype symmetry-protected topological insulator, i.e., the SSH chain \cite{ssh79},  described by a one-dimensional lattice Hamiltonian of spinless fermions with staggered hopping parameters, as illustrated in Fig.~\ref{fig:models}(a). There are two atoms in each primitive cell, labeled by $A$ and $B$. 
%The model can be regarded as consisting two sub-lattices of the atom $A$ and the atom $B$. 

\begin{figure}[t]
\centering
\begin{minipage}{\linewidth}
\centering
\includegraphics[width=\linewidth]{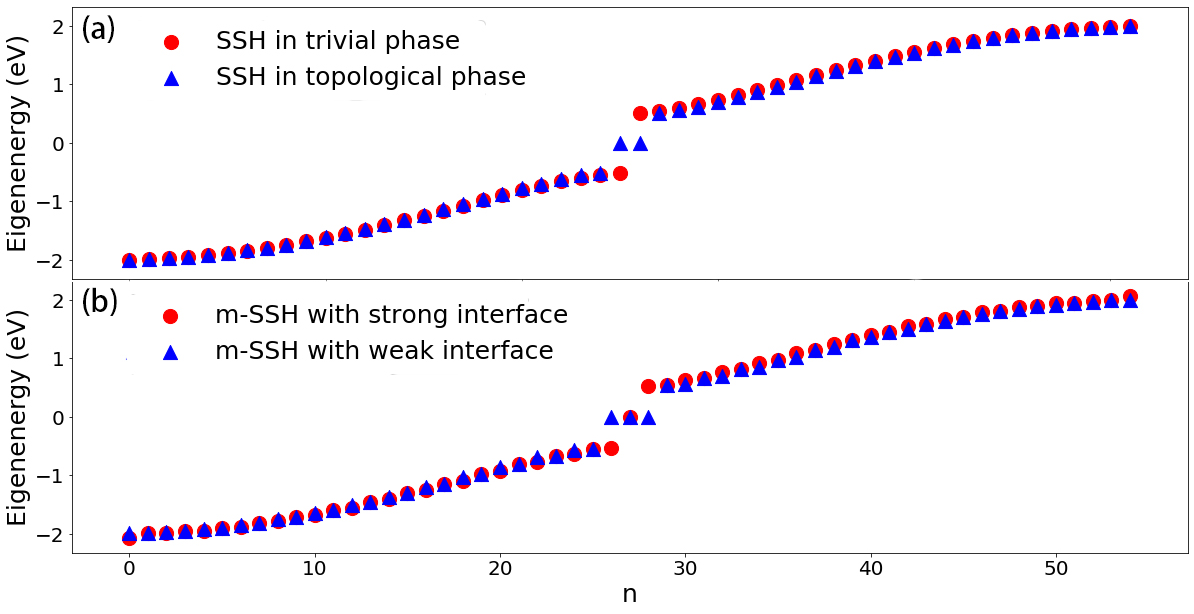}
\end{minipage}
\caption{(a) Energy spectrum of the SSH model on a 52-site open-ended chain. There are two zero-energy edge states in the topologically non-trivial sector ($t_1 = 0.75\text{ eV} < 1.25\text{ eV} = t_2$), but no edge state in the trivial sector ($t_1 = 1.25\text{ eV} > 0.75\text{ eV} = t_2$). (b) Energy spectrum  of the mirror SSH model on a 55-site open-ended chain. There is one zero-energy localized edge state for a strongly coupled inversion center and three zero-energy localized states for a weakly coupled inversion center.}
\label{fig:Spectrum}
\end{figure}

The real-space Hamiltonian of the SSH model is given by
\begin{eqnarray}
\hat{H} &=& t_1 \sum_{m=1}^{N}(\ket{m,B}\bra{m,A}+H.c.) \nonumber \\ &+& t_2 \sum_{m=1}^{N-1}(\ket{m+1,A}\bra{m,B}+H.c.) \nonumber \\
&+& \mu \sum_{m=1}^{N}(\ket{m,A}\bra{m,A} + \ket{m,B}\bra{m,B}),
\end{eqnarray}
where $N$ is the number of unit cells. $t_1$ and $t_2$ are the intra-cell and inter-cell hopping parameters, respectively. The open-ended SSH chain has two distinct topological sectors whose topological invariants can be represented by the number of zero-energy single-particle edge states $N_{es}$ in the energy gap. In Fig.~\ref{fig:Spectrum}(a), we show the electronic energy spectra of the SSH chain with 52 sites in both the topologically non-trivial sector ($t_1 = 0.75\text{ eV} < 1.25 \text{ eV} = t_2$) and in the trivial sector ($t_1 = 1.25 \text{ eV} > 0.75 \text{ eV} = t_2$). In the former case we observe two zero-energy edge states in the gap, whereas in the latter case there are no  edge states. Due to the bulk-boundary correspondence, these topological properties can also be identified via the bulk winding number $W$.

\begin{figure}[t]
\centering
\begin{minipage}{\linewidth}
\centering
\includegraphics[width=\linewidth]{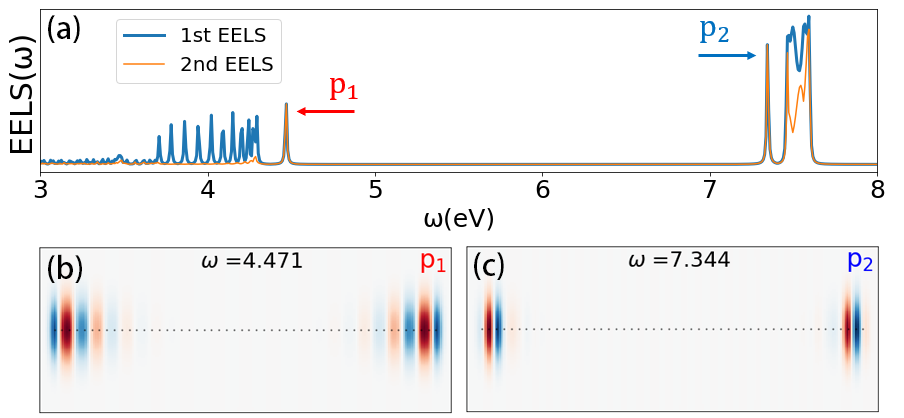}
\end{minipage}
\caption{(a) EELS of the 52-site open-ended SSH model in the topologically non-trivial sector ($t_1 = 0.75\text{ eV} < 1.25\text{ eV} = t_2$). (b) and (c): charge density modulations of the two localized plasmons in the first EELS at an intermediate and at a high frequency, only observed in the topological sector.}
\label{fig:SSH_full}
\end{figure}

We now focus on the SSH model in the topologically non-trivial sector ($t_1=0.75 \text{ eV and } t_2=1.25\text{ eV}$) and analyze its  plasmonic excitations. The EELS is shown in Fig.~\ref{fig:SSH_full}, where we observe two continua (mainly consisting of bulk modes) separated by an energy gap, as well as two isolated modes in the gap at $\omega=4.471 \text{ eV}$ and $\omega=7.344 \text{ eV}$, labeled by $p_1$ and $p_2$. The real-space charge density modulations of these two modes are shown in Figs.~\ref{fig:SSH_full}(b) and (c). We observe that they are both localized at the ends of the chain. Furthermore, the charge distribution of the higher frequency mode [Fig.~\ref{fig:SSH_full}(c)] is more strongly localized than the lower frequency mode [Fig.~\ref{fig:SSH_full}(d)]. We also point out that these two modes are both two-fold degenerate in the pure SSH model, as indicated by the peaks in both the 1st EELS and the 2nd EELS. Below, we will further study the effects of molecular perturbations on mode degeneracy.

\begin{figure}[t]
\centering
\begin{minipage}{\linewidth}
\centering
\includegraphics[width=\linewidth]{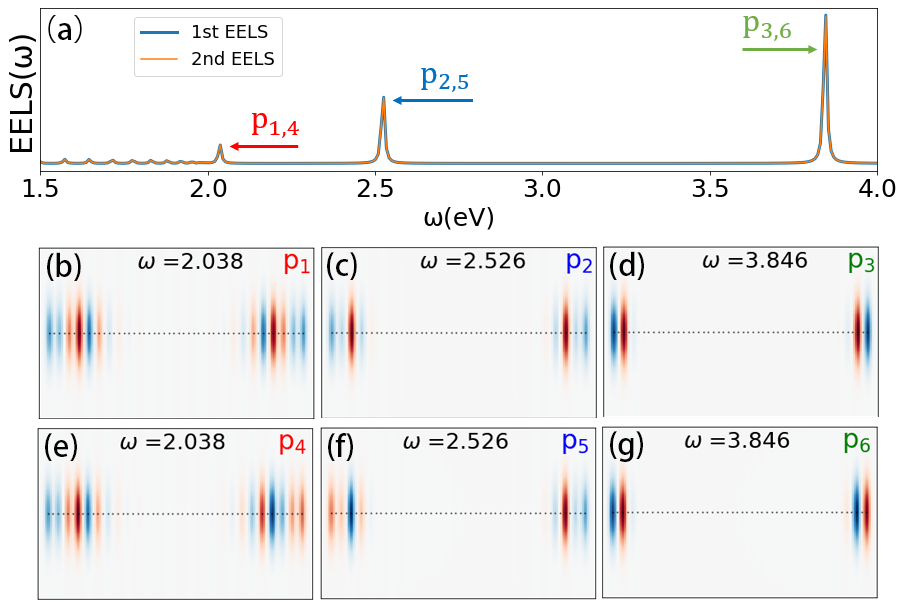}
\end{minipage}
\caption{(a) EELS using only the topological charge susceptibility $\chi_0^{\text{topo}}$ in the topological sector ($t_1 = 0.75\text{ eV} < 1.25\text{ eV} = t_2$) of the SSH model on an open-ended 52-site chain. (b) to (g): charge density modulations of the three degenerate localized plasmonic excitations at different frequencies. The edge modes observed in $\chi_0^{\text{full}}$ are preserved when only $\chi_0^{\text{topo}}$ is considered.}
\label{fig:SSH_Xtopo}
\end{figure}

In previous work, localized plasmons in open-ended TIs have been shown to originate from the topological electronic edge states \cite{jrgh20}. This was demonstrated by decomposing the full charge susceptibility $\chi_0^{\text{full}}$ into bulk and topological surface contributions, namely,
\begin{equation}
    \underbrace{\sum_{i,j}\dots}_{\chi_0^{\text{full}}} = \underbrace{\sum_{i\in\text{TS}}\,\sum_{j\notin\text{TS}}\dots+ \sum_{i\notin\text{TS}}\,\sum_{j\in\text{TS}}}_{\chi_0^{\text{topo}}}\dots + \underbrace{\sum_{i,j\notin\text{TS}}\dots}_{\chi_0^{\text{bulk}}},
\end{equation}
where TS is the set of the topological zero-energy edge states in the bulk gap. The spectrum of $\chi_0^{\text{topo}}$ preserves the plasmonic edge modes observed before in the full EELS spectrum, along with their degeneracies and their localized character in the real space. (see Fig.~\ref{fig:SSH_full} and Fig.~\ref{fig:SSH_Xtopo}). This allows us to focus on $\chi_0^{\text{topo}}$ instead of $\chi_0^{\text{full}}$ for an isolated examination of these localized edge plasmon modes. Below we calculate the EELS of the SSH model, using only $\chi_0^{\text{topo}}$, and denote the resulting spectrum by $\text{EELS}^\text{topo}(\omega)$, which is shown in Fig.~\ref{fig:SSH_Xtopo}(a). As expected, there is no bulk plasmonic continuum in the spectrum because of removal of the bulk contributions, $\chi_0^\text{bulk}$.  $\text{EELS}^\text{topo}(\omega)$ shows three  peaks at $\omega=2.038\,\text{ eV},\,2.526\,\text{ eV} \,\text{and} \,3.846\,\text{ eV}$. Each mode has a two-fold degeneracy of different parities: odd parity and even parity, and they all have localized charge distributions [see Figs.~\ref{fig:SSH_Xtopo}(b) to (g), up: even parity; down: odd parity]. Also, similar to the $\text{EELS}^\text{full}(\omega)$ with full susceptibility, the edge plasmons are more strongly localized at higher frequencies,  resembling the patterns shown in Fig.~\ref{fig:SSH_full}. Furthermore, note that even though the bulk susceptibility does not contribute to the localization of the edge plasmons, it still affects the excitation energies of these modes.  

\begin{figure}[t]
\centering
\begin{minipage}{\linewidth}
\centering
\includegraphics[width=\linewidth]{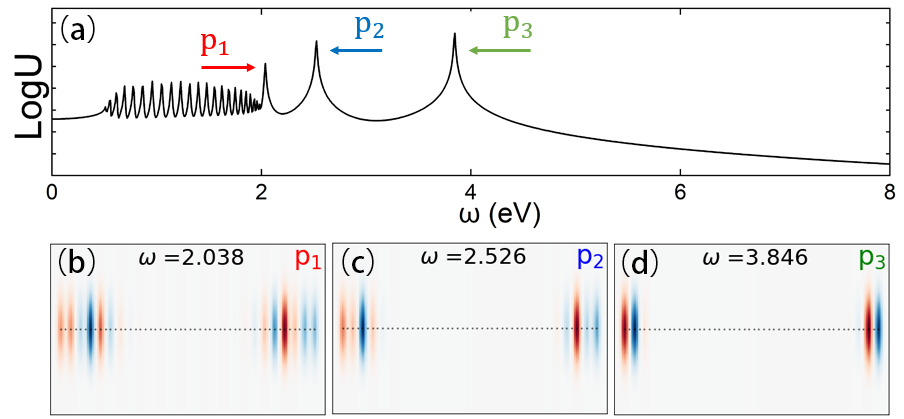}
\end{minipage}
\caption{(a) Induced energy spectrum in the 52-site open-ended SSH chain in the topological sector ($t_1 = 0.75\text{ eV} < 1.25 eV = t_2$), subject to a linear external electrical field. (b), (c) and (d): charge density modulation of the mode corresponding to the three  excitations highlighted in (a).}
\label{fig:SSH_Ex}
\end{figure}

 We also calculate the induced energy spectrum $U_{\text{ind}}(\omega)$ of the SSH model in response to a specific external electromagnetic field, which can be directly compared to experiments. Here we consider a linear external electric potential applied to a finite SSH chain in the topologically non-trivial sector. The resulting spectrum shown in Fig.~\ref{fig:SSH_Ex}(a) contains three main peaks at the exactly same frequencies as those obtained from $\text{EELS}^\text{topo}(\omega)$ [Fig.~\ref{fig:SSH_Xtopo}(a)], which confirms the eigen-modes indicated EELS. However, The induced charge distributions of these modes in Figs.~\ref{fig:SSH_Ex}(b)-(d) are odd functions in the real space, whereas their even-parity partners from the EELS [Figs.~\ref{fig:SSH_Xtopo}(b)-(d)] become inactive now. This is expected because, under the linear potential, only modes with odd parity are excited.

\subsection{Effects of added diatomic molecules on plasmons in the SSH chain}\label{sec:3.3}
The topological SSH model studied above hosts both bulk plasmons and localized plasmonic edge modes \cite{jrgh20}. While the former show no essential difference from the propagating modes in the 1D MC in Sec.~\ref{sec:A}, the localized edge plasmons respond differently to molecular perturbations. Here we introduce diatomic molecules in the vicinity of the SSH chain and study their effects on the plasmonic excitations. Specifically, we place a single diatomic molecule above the SSH chain and gradually change its position from the edge to the center of the chain, as illustrated in Fig.~\ref{fig:models}(c). The tunneling hopping between the molecule and the SSH chain is denoted by $\tilde{t}$, with $\tilde{t}<t_1,t_2$). The internal hopping in the molecule is denoted as $t'$, with $t'>t_1,t_2$.  Such a situation could be experimentally realized by atoms attached on an STM tip or by tip atoms themselves when scanning over a sample, such as described in ~\cite{fl03}. Here, we focus on studying the effects of the perturbing molecule at various positions on the plasmonic edge modes.  

\begin{figure}[t]
\centering
\begin{minipage}{\linewidth}
\centering
\includegraphics[width=\linewidth]{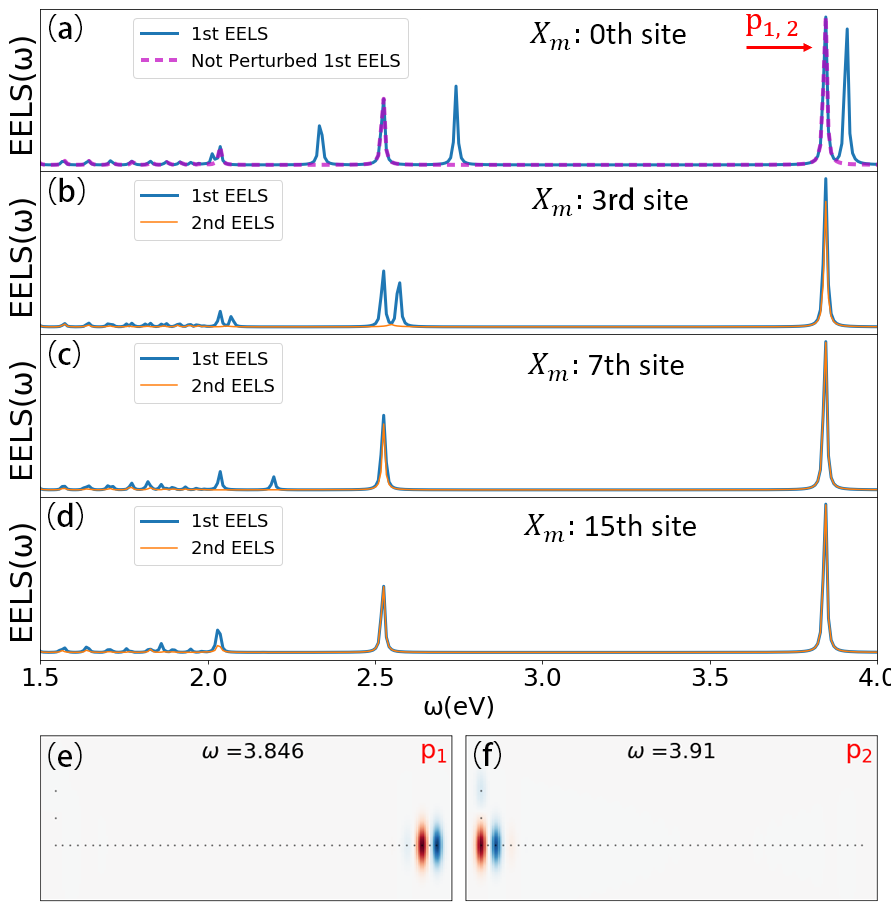}
\end{minipage}
\caption{(a) to (d): Topological EELSs in the SSH chain ($t_1 = 0.75\text{ eV} < 1.25\text{ eV} = t_2$) with one added diatomic molecule at different positions, $X_m$,  which means the molecule is connected with the $x^{\text{th}}$ site on the chain like Fig. \ref{fig:models}(c). The connection hopping $\tilde{t}=0.5\text{ eV}$ and $t'=2.0\text{ eV}$. (e) and (f): charge density modulation of the modes corresponding to the two high-frequency excitations in (a).}
\label{fig:SSH_per}
\end{figure}

We first consider the effect of a diatomic molecule in proximity to one of the ends of the SSH chain [Fig.~\ref{fig:models}(c)], which is expected to have  maximum impact on the plasmonic edge modes. Fig.~\ref{fig:SSH_per}(a) shows the EELS of such a perturbed SSH chain, together with the unperturbed case for comparison. 
We  find that the molecular perturbation on the edge site removes the degeneracy of all three modes observed in the pure host system. Due to the additional molecule  attached to one edge site, the two ends of the chain are no longer equivalent. Therefore, they now each host edge modes with slightly different energies. For instance, in Figs.~\ref{fig:SSH_per}(e) and (f) we show the real-space charge distribution patterns for the two highest energy modes [labeled as $p_1$ and $p_2$ in Fig.~\ref{fig:SSH_per}(a)]. They originate from a degenerate pair at $\omega=3.486\ \text{eV}$ of the host model. In the presence of the molecular perturbation applied to the left end of the open chain, the mode localized on the right end of the chain remains at the same frequency as before because of its far distance away from the perturbing molecule, whereas the edge mode at the left end of the chain now has a slightly shifted energy. Naturally, plasmons that are localized close to the left end of the chain are mostly affected by this local perturbation.

As we gradually move the perturbing molecule from the left end to the center of the chain, the effects on the plasmonic edge modes become less pronounced. Quantitatively, this  depends on the localization length of the plasmonic edge modes. As mentioned above, the highest-energy mode is most localized. The two lower-energy modes are slightly more extended [Figs.~\ref{fig:SSH_Xtopo}(b)-(g)]. When the molecule is being moved towards the center of the chain, the highest-energy mode first becomes unaffected to the perturbation, then followed by the lower-energy ones. In Figs.~\ref{fig:SSH_per} (b)-(d) we show the variation of the EELS, as the molecular perturbation is gradually moved to the center. In detail, we can see that when the perturbing molecule is moved onto the $3^\text{rd}$ site away from the the chain edge, the highest-energy mode is already not affected. However, the two lower-energy modes are still affected by the perturbation, as we can see from the split peaks. When the perturbing molecule is on the $7^\text{th}$ site of the chain, the second-highest-energy mode becomes unaffected as well [Fig.~\ref{fig:SSH_per}(c)]. Finally, when the molecule is on the $15^\text{th}$ site of the chain, which is quite deep into the bulk, all three modes are unaffected. In this case, the full EELS is almost the same as for the unperturbed host system. 

\begin{figure}[t]
\centering
\begin{minipage}{\linewidth}
\centering
\includegraphics[width=0.75\linewidth]{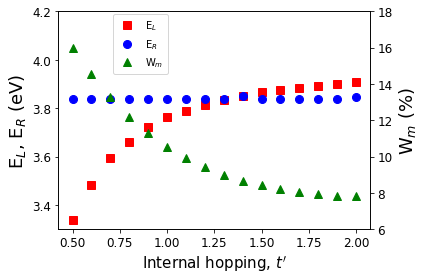}
\end{minipage}
\caption{Dependence of the energy of the topological plasmons and of the charge transferred from the host to the diatomic molecule on the internal hopping $t^\prime$, when the molecule is connected with an edge atom of the topological SSH chain ($t_1 = 0.75\text{ eV} < 1.25\text{ eV} = t_2$). $E_L$ (Red) is the highest energy excitation on the left end of the chain (see Fig.~\ref{fig:SSH_per}(f)). $E_R$ (Blue) is the highest energy excitation on the right end of the chain (see Fig.~\ref{fig:SSH_per}(e)). $W_m$ is the percentage weight of the charge transferred from the chain to the molecule.}
\label{fig:Diff_t}
\end{figure}

 In the above calculations, the internal hopping $t^\prime$ was fixed to 2.0 eV, which is larger than the hopping in the chain. Here, we would like to examine how the the excitation energies change when we modify $t^\prime$. When we consider only the topological susceptibility, the excitations  are localized at the two ends of the chain. In  Fig.~\ref{fig:Diff_t} we see that the excitation on the right end remains constant at about $E_R =$ 3.84 eV when the perturbing molecule is connected to the left end. However, the excitation at the left end, $E_L$, shifts  to higher energies when $t^\prime$ is increased because a higher energy is required to excite the molecule with a larger internal energy gap $t^\prime$. Furthermore, we find that the charge  transferred from the chain to the molecule also changes with $t^\prime$. $W_m$ in Fig.~\ref{fig:Diff_t} shows the percentage weight of the charge on the molecule compared to the charge in the entire system (host chain plus perturbation molecule). We observe that the relative weight on the molecule drops from about 16\% to about 8\% when $t^\prime$  is increased, i.e. there is less charge transfer from the host to the molecule. Hence, the internal electronic structure of the added molecule affects its ability to hybridize with the host system. When the internal hopping ($t_1$ and $t_2$) inside the SSH chain is fixed, if the molecule has a larger internal energy gap, i.e. larger $t^\prime$, then it is more protected from hybridization with the host system.

We conclude that a local perturbation can affect the topologically originated plasmonic edge modes in the SSH model only significantly when it is sufficiently close  to the edge of the chain. In other words, the edge plasmon modes in the topological SSH model are very robust against local perturbations, which is different from bulk plasmon modes. It has been shown before that these modes are also quite stable when subjected to global random noise in the bulk hopping parameters\cite{jrgh20}.
\label{sec:C}

\subsection{Plasmonic excitations in the mirror-SSH model}\label{sec:3.4}
In addition to the SSH model, let us also inspect plasmons in the mirror-SSH model, which is a variant of the SSH model by reflecting the chain about its center [Fig.~\ref{fig:models}(b)]. This mirror-SSH (mSSH) model is inversion symmetric with respect to the mirror interface located at the middle point of the chain, which also hosts localized zero-energy state(s) depending on the hopping characteristics at the interface. In Fig.~\ref{fig:Spectrum}(b), the energy spectra of the strong interface mSSH model ($t$ at the center is 1.25 eV) and the weak interface mSSH model ($t$ at the center is 0.75 eV) are displayed, where we observe one zero-energy state and three zero-energy states, respectively.

\begin{figure}[t]
\centering
\begin{minipage}{\linewidth}
\centering
\includegraphics[width=0.95\linewidth]{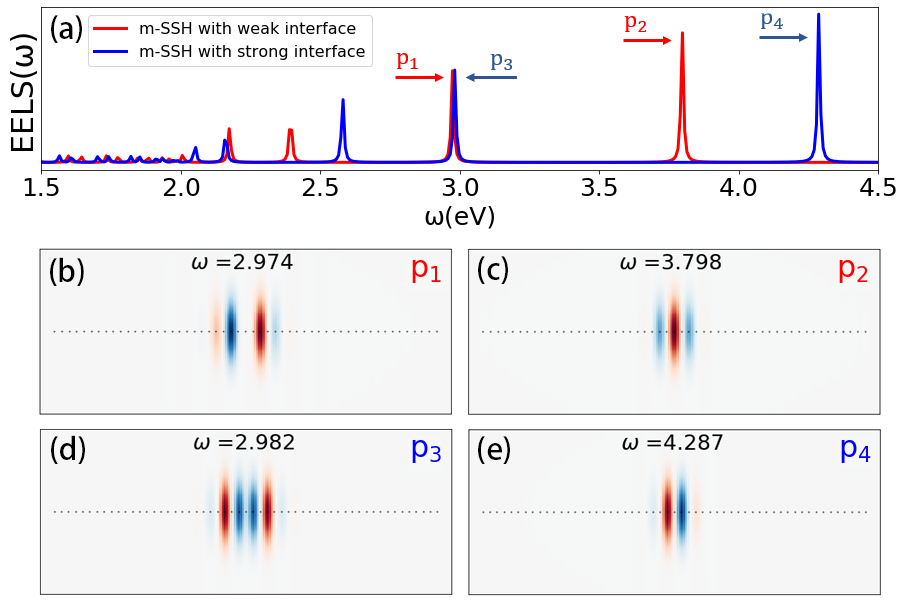}
\end{minipage}
\caption{(a) EELS of the open-ended mirror-SSH chain (M=53) with weak mirror interface hopping ($t_1 = 0.75 eV < 1.25 eV = t_2$) and on an open-ended 55-site chain with strong mirror interface hopping ($t_1 = 1.25 eV >0.75 eV = t_2$). (b) and (c): charge density modulation of the modes corresponding to the excitations at the weak mirror interface. (d) and (e): charge density modulation of the modes corresponding to the excitations at the strong mirror interface.}
\label{fig:mSSH_EELS}
\end{figure}
 
 We calculate the EELS of the mSSH model plasmons by using only the ``topological part" of the susceptibility $\chi_0^{\text{topo}}$, defined in the same manner as for the SSH model. Fig.~\ref{fig:mSSH_EELS}(a) shows the EELS of the mSSH model for both the strong mirror interface (hopping at the mirror is $t_1 = 1.25\text{ eV}$) and the weak mirror interface (hopping at the mirror is $t_2 = 0.75\text{ eV}$). In each case, we observe a set of (non-degenerate) excitations that are all strongly localized around the mirror interface. We show the real-space charge modulations of some typical excitations in Figs.~\ref{fig:mSSH_EELS} (b)-(e). As we can see here, the modes around the strong interface and the weak interface have different parities. Similar to what we observed before in the SSH model, the modes with higher energies are more strongly localized.

\subsection{Effects of added diatomic molecules in the mirror-SSH chain}\label{sec:3.5}
Here we introduce a single molecular perturbation into the mSSH model, starting from the central site (the mirror interface) of the chain and moving towards one end, as shown in Fig.~\ref{fig:models}(d). We focus on studying its effects on the localized plasmons around the interface. Unlike the edge plasmons observed in the SSH model, the interface plasmons here are non-degenerate even in the unperturbed case. So, there is no degeneracy splitting effect. We will, however, observe other interesting effects due to the perturbation.

\begin{figure}[t]
\centering
\begin{minipage}{\linewidth}
\centering
\includegraphics[width=0.95\linewidth]{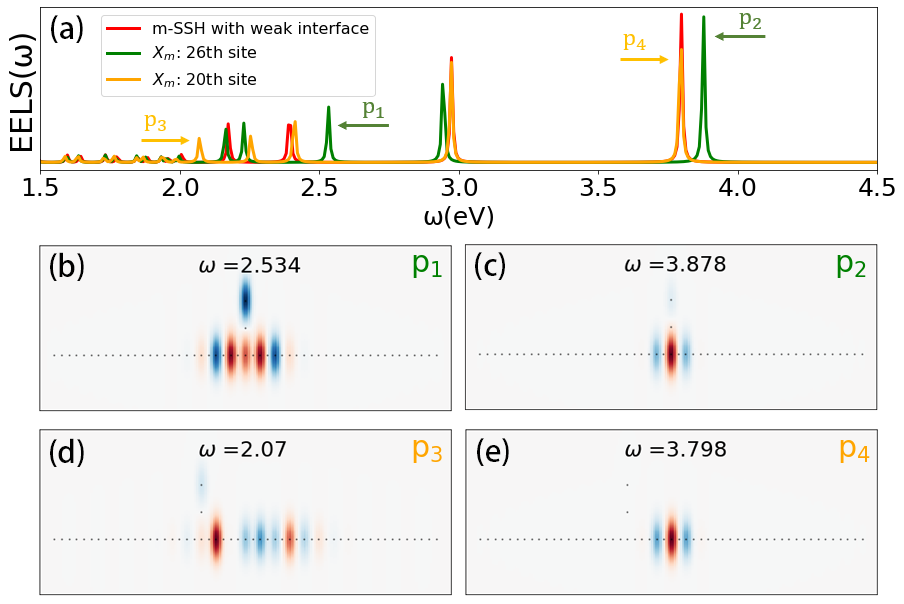}
\end{minipage}
\caption{(a) Comparison of the EELS of the weak interface mirror-SSH model ($t_1 = 0.75\text{ eV} < 1.25\text{ eV} = t_2$) with M=53 sites in the presence of a perturbation diatomic molecule placed at different positions (green line:  connected with the $26^{\text{th}}$ site; orange line:  connected with the $20^{\text{th}}$ site) and the model without perturbation (red line). (b) - (e): charge density modulations of the modes corresponding to the excitations highlighted in (a).}
\label{fig:mSSH_weak_per}
\end{figure}

Fig.~\ref{fig:mSSH_weak_per}(a) shows the perturbed EELS of the weak-interface mSSH model, together with the unperturbed one (red line) for comparison. The spectra are similar to the SSH model discussed in section (c), in the sense that  the effects from the perturbation are weakened when the perturbing molecule is gradually shifted away from the charge concentration area of the localized plasmons. When the added molecule is connected to the $26^{\text{th}}$ site, which is the center site of the 53-site chain, all of the excitations change their positions (green line) because of significant charge transfer between the chain and the molecule. Figs.~\ref{fig:mSSH_weak_per}(b) and (c) show the charge distributions of two typical excitations in (a). Here, we observe that there is little charge transfer from the chain to the molecule in the high energy excitation, but more significant charge transfer to the molecule connected to the center site for the lower energy mode. The reason for this is that the internal hopping magnitude of the molecule is $t'=2\ \text{eV}$, which is closer to the frequency of the low energy excitation. However, when we move the perturbation from the center towards the edge, the EELS curve will coincide with the original curve (red line) in the high and intermediate energy regimes, since the charge is more concentrated at the center in the higher energy modes [Fig.~\ref{fig:mSSH_weak_per}(e)], so the interaction between the chain and the molecule substantially vanish. In this case, there is still charge transfer in lower energy modes, and the charge distribution will not be symmetrical anymore [Fig.~\ref{fig:mSSH_weak_per}(d)]. 

\begin{figure}[t]
\centering
\begin{minipage}{\linewidth}
\centering
\includegraphics[width=0.95\linewidth]{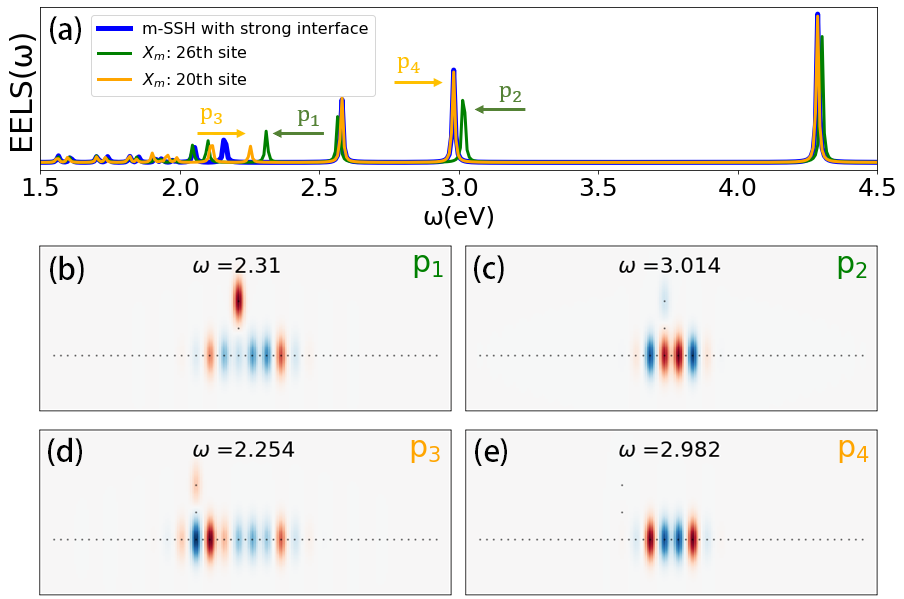}
\end{minipage}
\caption{(a) Comparison of the EELS of the strong interface mirror-SSH model ($t_1 = 1.25\text{ eV} > 0.75\text{ eV} = t_2$) in the presence of a diatomic perturbation molecule placed at different positions (green line: connected with the $26^{\text{th}}$ site; orange line: connected with the $20^{\text{th}}$ site) and the model without perturbation (blue line). (b) - (e): charge density modulations of the modes corresponding to the excitations indicated in (a).}
\label{fig:mSSH_strong_per}
\end{figure}
Next, we discuss the strong interface case, whereby the hopping at the center mirror interface is stronger in  magnitude. Fig.~\ref{fig:mSSH_strong_per}(a) shows a comparison of the strong interface with (red and orange line) or without (blue line) the molecular perturbation. For the strong interface, since the plasmon is also concentrated in the center we observe analogous spectra as for the weak interface. Specifically, the effect of the perturbation molecule on topological surface plasmons in the mirror-chain decreases when we move it from the chain center towards the chain ends, especially in the higher energy regime. When $\omega$ is larger than $2.5\text{ eV}$, the blue curve and orange curve coincide with each other, which means there is little effect when the molecular perturbation is connected with the $20th$ site than connected with the $26^{\text{th}}$ site. In the low energy region, however,  we  also observe an excitation shift caused by charge transfer (Fig.~\ref{fig:mSSH_strong_per}(d)) because the plasmonic excitation expands to the $20^{\text{th}}$ site of the chain. However, when the molecular perturbation is connected with the $26^{\text{th}}$ of the chain (in the center), all of the excitations of the host system shift from low energy  to high energy, and  plasmonic charge is transferred to the molecule [Figs.~\ref{fig:mSSH_strong_per}(b) and (c)].

\section{Conclusions}\label{sec:4}
In this work, we have examined the effects of added diatomic molecules on the plasmonic modes of 1D  SSH model and its mirrored invariant,  comparing them to the benchmark of a simple metallic chain. By analysis of the electron energy loss excitation spectrum and of the real space charge distributions of the plasmon modes, we conclude that the position of the local perturbation is the key parameter to control the plasmons in 1D TIs. When the perturbation is on or near the edges of the topological insulator, the plasmonic excitations in the topologically non-trivial regime, i.e., their degeneracy and their charge distribution, will be significantly affected. In contrast, the plasmonic excitations become less affected when the local perturbation is far from the edges. We also identified conditions under which charge transfers from the host chain to the added molecules. Here, the internal hopping $t^\prime$ within the diatomic molecule plays an important role. It will be interesting to further analyze analogous effects of perturbing molecules in higher-dimensional topological systems that harbor dispersive surface modes, such as the two-dimensional SSH model and graphene, which may become localized due to the impurities.

\begin{acknowledgments}
We wish to acknowledge useful discussions with Henning Schl{\"o}mer and Malte R{\"o}sner. This work was supported by the US Department of Energy under grant
number DE-FG02-05ER46240.
\end{acknowledgments}

\appendix
\section{Induced field energy of the perturbed SSH model}

\begin{figure}[t]
\centering
\begin{minipage}{\linewidth}
\centering
\includegraphics[width=\linewidth]{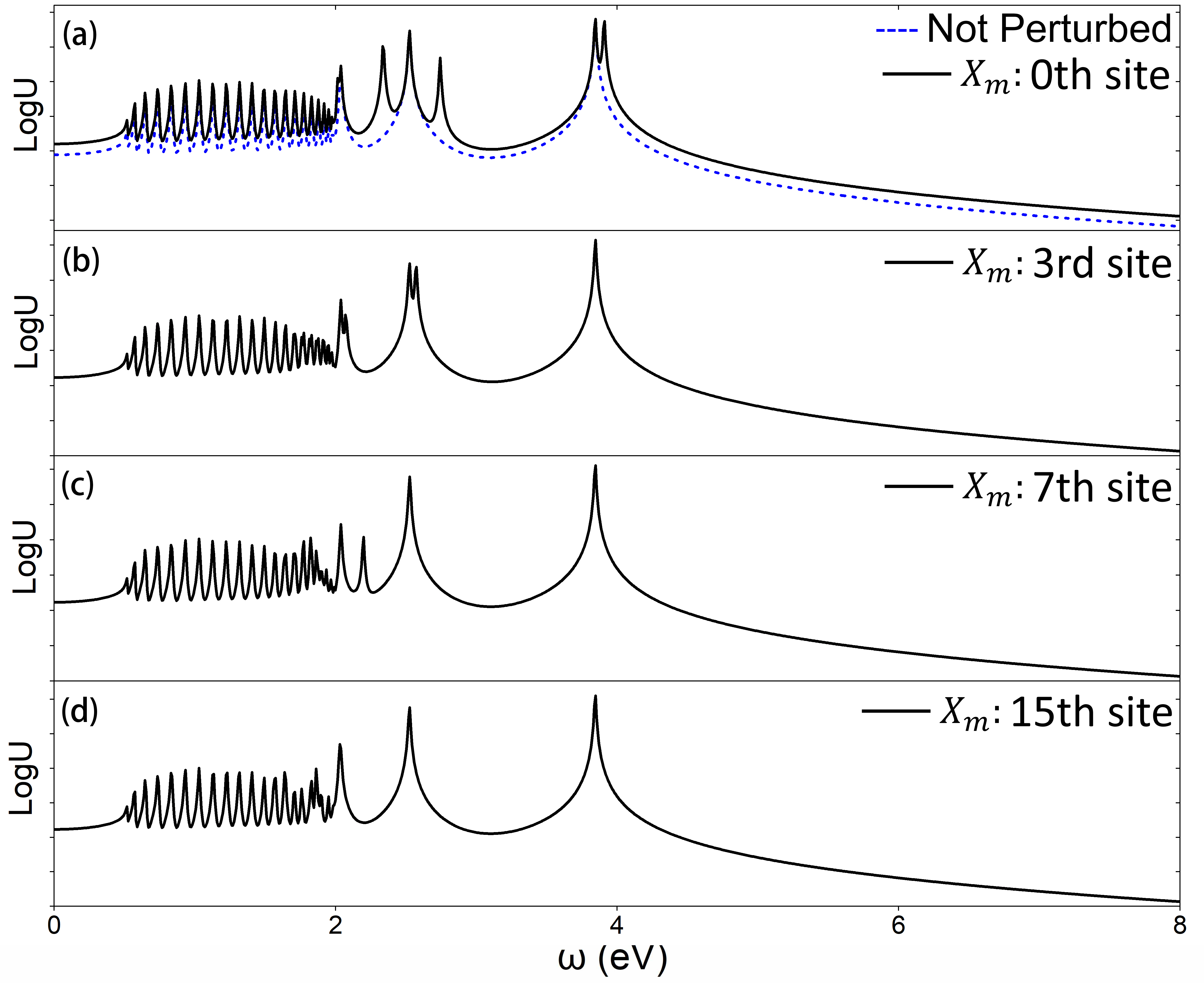}
\end{minipage}
\caption{Energy spectrum, induced by a linear external electrical field, of a 52-site non-trivial topological SSH chain ($t_1 = 0.75\text{ eV} < 1.25\text{ eV} = t_2$) with one diatomic perturbation molecule located at different positions, calculated using only the topological susceptibility $\chi_0^{\text{topo}}$.}
\label{fig:SSH_External_per}
\end{figure}

In Sec.~\ref{sec:3.2}, we discussed the induced energy spectrum of the SSH chain and analyzed the spectrum in Fig.~\ref{fig:SSH_Ex}. We observe three excitations at  exactly the same positions as in Fig.~\ref{fig:SSH_Xtopo}. In Sec.~\ref{sec:3.3}, the effects of local perturbation were considered. Here, we analyze the induced energy spectrum of the topological SSH model in the presence of a diatomic molecular perturbation, subject to an external linear electrical field. As a reference, the blue dashed line in Fig.~\ref{fig:SSH_External_per}(a) shows the  spectrum without perturbation that we have already discussed. Figs.~\ref{fig:SSH_External_per} (a)-(d) reveal that the effects of the perturbation to SSH model in the presence of an external electrical field are similar to the EELS spectrum, i.e.,  mostly affecting the degeneracy with respect to the position. As the perturbing molecule is gradually moved from the left end ($0^{\text{th}}$ side) towards the center of the bulk ($15^{\text{th}}$ site), the effects on the plasmonic edge modes become less pronounced. The highest-energy excitation first becomes unaffected to the perturbation, as seen in Fig.~\ref{fig:SSH_External_per}(b), and then the lower energy modes recover. When the perturbation moves sufficiently deep into the bulk, the energy spectrum recovers to the same shape of the non-perturbed system.

\nocite{*}

\bibliography{ref}% Produces the bibliography via BibTeX.

\end{document}